\documentclass{article}
\usepackage{spconf}

\usepackage{graphicx}
\usepackage{amsmath}
\usepackage{amsfonts}
\usepackage{amssymb}
\usepackage[numbers,sort&compress]{natbib}
\usepackage[utf8]{inputenc}
\usepackage[T1]{fontenc}
\usepackage{booktabs}
\usepackage{makecell}
\usepackage{arydshln}
\usepackage{hyperref}
\usepackage{array}
\usepackage{float}
\title{Adapting HFMCA to Graph Data: Self-Supervised Learning for Generalizable fMRI Representations}

\name{Jakub Frąc$^1$ \quad Alexander Schmatz$^1$ \quad Qiang Li$^2$ \quad
Guido Van Wingen$^3$ \quad Shujian Yu$^1$}

\address{$^1$VU Amsterdam \quad $^2$TReNDS Center \quad $^3$Amsterdam UMC }

\begin{document}

\maketitle

\begin{abstract}

Functional magnetic resonance imaging (fMRI) analysis faces significant challenges due to limited dataset sizes and domain variability between studies. Traditional self-supervised learning methods inspired by computer vision often rely on positive and negative sample pairs, which can be problematic for neuroimaging data where defining appropriate contrasts is non-trivial. We propose adapting a recently developed Hierarchical Functional Maximal Correlation Algorithm (HFMCA) to graph-structured fMRI data, providing a theoretically grounded approach that measures statistical dependence via density ratio decomposition in a reproducing kernel Hilbert space (RKHS),
and applies HFMCA-based pretraining to learn robust and generalizable representations. Evaluations across five neuroimaging
datasets demonstrate that our adapted method produces competitive embeddings for various classification tasks and enables effective knowledge transfer to unseen datasets.
Codebase and supplementary material can be found here: \url{https://github.com/fr30/mri-eigenencoder}

\end{abstract}


\section{Introduction}

Functional magnetic resonance imaging (fMRI) provides crucial insights into human brain dynamics, with resting-state functional connectivity serving as an important biomarker for neurological and psychiatric conditions \citep{autism1,mdd1,schizophrenia1}. Yet, deep learning applications face major challenges due to limited dataset sizes, heterogeneous preprocessing protocols, and persistent domain shifts across scanning centers.


Contrastive self-supervised learning (SSL) has offered promising solutions by adapting approaches from computer vision to neuroimaging data \citep{fmrissl1,fmrissl2,fmrissl3,ss_transformers}. While some methods operate directly on temporal blood oxygen level-dependent (BOLD) signals \citep{fmrissl1, fmrissl3}, graph-based approaches that model functional connectivity matrices offer distinct advantages for neuroimaging applications. Functional connectivity graphs provide a more structured and interpretable representation of brain organisation, capturing pairwise statistical dependencies between brain regions whilst reducing the dimensionality and complexity of the raw temporal signal space. Additionally, graph-based augmentations such as node sampling and edge perturbation preserve the underlying network topology.


A recently proposed \textbf{H}ierarchical \textbf{F}unctional \textbf{M}aximal \textbf{C}orrelation \textbf{A}lgorithm (HFMCA) \citep{hfmca}, originally used for image data, provides a theoretically principled approach with strong potential to address these neuroimaging-specific challenges. Unlike contrastive methods that rely on explicit positive–negative sample construction \citep{simclr, moco}, HFMCA measures statistical dependence between low- and high-level features across multiple views without being limited to traditional two-view frameworks \citep{barlowtwins, vicreg}. This multi-view capability enables the capture of richer hierarchical dependencies, leading to greater feature diversity and more generalizable representations. HFMCA operates on graph-structured connectivity rather than raw BOLD signals, exploiting brain network topology to integrate complementary views of neural activity, making it well-suited for neuroimaging representation learning.



\textbf{Contributions:}
    (1) We adapt HFMCA to graph-structured fMRI data, representing the first application and extension of this framework to brain connectivity networks. 
    (2) We demonstrate that HFMCA-pretrained encoders produce competitive embeddings for neuroimaging classification tasks across diverse datasets. 
    (3) We show effective transfer learning capabilities, particularly in scenarios where limited labelled data is available.
    4) We evaluate neural scaling laws in the context of fMRI graph encoders, showing that naive pretraining data scaling may induce negative transfer.
    
\section{Background}
\subsection{Functional Maximal Correlation Algorithm}

The Functional Maximal Correlation Algorithm \cite{hfmca} measures statistical dependence between random variables $X$ and $Y$ through their probability distributions. For distributions $p(X)$ and $p(Y)$, statistical dependence is characterised by:
\begin{equation}
\rho(X,Y) := \frac{p(X,Y)}{p(X)p(Y)}.
\end{equation}
FMCA approximates this dependence by decomposing it into eigenvalues $\sigma_k$ and corresponding orthogonal eigenfunctions:

\begin{gather}
\rho(X,Y) = \frac{p(X,Y)}{p(X)p(Y)} \approx \sum_k\sqrt{\sigma_k}\varphi_k(X)\psi_k(Y) \\
\mathbb{E}_X[\varphi_k(X)\varphi_{k^\prime}(X)] = \mathbb{E}_X[\psi_k(Y)\psi_{k^\prime}(Y)] = \begin{cases}
    1,\text{ }k = k^\prime \\
    0,\text{ }k \neq k^\prime
  \end{cases}
\end{gather}

Two neural network encoders $f_{\theta}: \mathcal{X} \rightarrow \mathbb{R}^K$ and $g_{\omega}: \mathcal{Y} \rightarrow \mathbb{R}^K$ approximate the eigenfunctions $\varphi$ and $\psi$ respectively. The autocorrelation matrices are computed as:
\begin{gather}
    R_F = \mathbb{E}_X[f_\theta(X)f_\theta^\top(X)], \quad R_G = \mathbb{E}_Y[g_\omega(Y)g_\omega^\top(Y)] \\
    P_{FG} = \mathbb{E}_{X,Y}[f_\theta(X)g_\omega^\top(Y)]
\end{gather}

The joint autocorrelation matrix combines these components:
\begin{equation}
R_{FG} = \begin{bmatrix}
R_F & P_{FG} \\
P_{FG}^\top & R_G
\end{bmatrix}
\end{equation}

The FMCA objective maximises statistical dependence by minimising:
\begin{equation}
\mathcal{L}_{FMCA} = \log \det R_{FG} - \log \det R_F - \log \det R_G
\end{equation}

This formulation encourages orthogonal feature learning (maximising $\log \det R_F$ and $\log \det R_G$) while aligning representations between views (minimising $\log \det R_{FG}$). 


Recently, FMCA has been employed in self-supervised image representation learning via hierarchical mutual information maximization \citep{hfmca}, and in cross-modal representation learning for EEG and EMG \citep{hfmca2}.
These applications demonstrate FMCA's capacity to capture statistical dependencies across different data modalities and feature hierarchies. We extend this framework to develop a self-supervised learning scheme for functional correlation matrix graphs, where the encoder learns semantically meaningful latent representations of brain connectivity patterns.


\section{Methods}

\subsection{Problem formulation}
Given a small labelled clinical dataset \( D_c = \{(X_i, Y_i)\}_{i=1}^{N_c} \) and a large population dataset \( D_p = \{X_j\}_{j=1}^{N_p} \) with \( N_p \gg N_c \), each subject is represented by a functional connectivity matrix \( X_i \in \mathbb{R}^{|V| \times |V|} \), computed over \( |V| \) atlas-defined brain regions. Each \( X_i \) encodes the pairwise statistical dependencies of regional BOLD signals and is modelled as a weighted graph on \( |V| \) nodes. The clinical label \( Y_i \in \{0,1\} \) indicates a binary diagnosis assigned by clinicians for a specific disorder. The objective is to learn a predictive model \( f: \mathbb{R}^{|V| \times |V|} \rightarrow \{0,1\} \) defined on the clinical dataset. Training \( f \) directly on \( D_c \) is susceptible to overfitting due to the limited sample size and variability in labels and data quality. 

To address this, we pretrain a neural encoder $f^1_\theta$ on \( D_p \) with a self-supervised objective to obtain semantically meaningful latent representations \( Z \) from connectivity graphs, which are then fine-tuned on \( D_c \) for diagnostic prediction. 

\subsection{Graph Construction from fMRI Data}

We preprocess resting-state fMRI recordings by parcellating each subject's brain into 116 anatomical regions using the AAL116 atlas \citep{aal116}. For each region of interest, we extract mean BOLD time series and compute Pearson correlation coefficients between all region pairs, yielding symmetric $116\times 116$ functional connectivity matrices. 
To build graphs for neural network processing, we retain the top $\tfrac{|V|^2}{400}$ correlation coefficients as edges, using connectivity values as node features. That is, for each network $G$, the feature of node $k$ is defined as $X_k = [\rho_{k1}, \ldots, \rho_{k|\mathcal{V}|}]^T \in \mathbb{R}^{|\mathcal{V}|}$, where $\rho_{kq}$ is the Pearson correlation between BOLD signals at nodes $k$ and $q$.


\begin{figure}[t]
\begin{minipage}[b]{1.0\linewidth}
  \centering
  \centerline{\includegraphics[width=8.5cm]{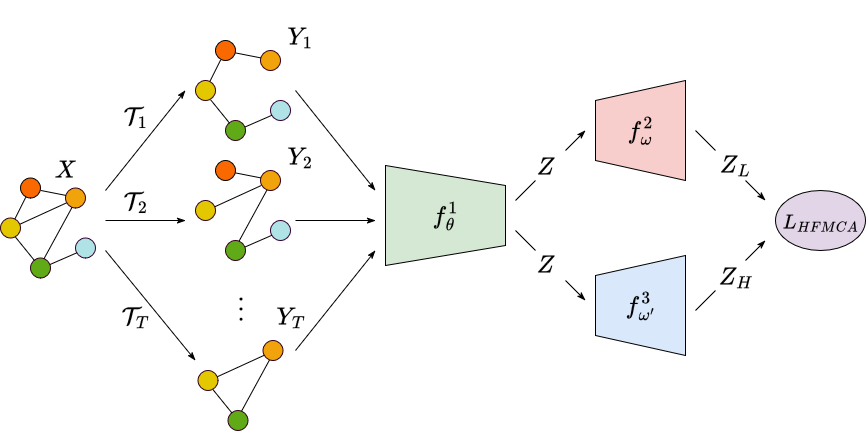}}
\end{minipage}
  \caption{HFMCA learns representations by maximising the statistical dependence between low- and high-level features of the graph. Each subsampling augmentation (e.g., random walk sampling) is processed through a shared backbone \( f^1_\theta \), producing low-level features. These features are either projected individually through a local head $Z^L = [f^2_\omega \circ f^1_\theta(Y_1) \dots f^2_\omega \circ f^1_\theta(Y_T)]$ or jointly aggregated through a high-level head $Z^H = \sum_i^Tf^3_{\omega'_i} \circ f^1_\theta(Y_i)$. In pretraining, both projection heads are used to enforce multi-view consistency, while only the backbone is retained for downstream tasks.}
  \label{fig:hfmca}
\end{figure}

\subsection{Hierarchical FMCA with Graphs}
Motivated by the original framework developed for image
classification \cite{hfmca}, we extend HFMCA to graph-structured fMRI data. HFMCA models dependencies between hierarchical feature representations of brain regions rather than pairwise correlations between different data views. Given a source graph $X$ and $T$ augmentations $Y = \{Y_1, Y_2 \ldots, Y_T\}$ where $Y_i = \mathcal{T}_i(X)$, the method defines:\\
\textbf{Low-level features:} $Z^L = \{Z^L_t\}^T_{t=1}$ from individual augmentation representations $Z$. \\
\textbf{High-level features:} $Z^H$ from aggregated augmentation representations $Z$.

The hierarchical dependence is modelled as:
\begin{equation}
\rho(Z^L, Z^H) = \frac{p(Z^L, Z^H)}{p(Z^L)p(Z^H)}.
\end{equation}


Three network components extract these features: backbone encoder $f_\theta^1$, and projection heads $f_\omega^2$ and $f_{\omega'}^3$ (Figure \ref{fig:hfmca}).
For each brain connectivity graph \( X \) with augmentations \( Y = \{Y_1, \ldots, Y_L\} \), where \( Y_i = T_i(X) \), the set of embeddings is defined as $Z = \{f^{1}_{\theta}(Y_1), \ldots, f^{1}_{\theta}(Y_T)\}$.
These features are passed through the projection heads:\\
\hfill \\
$\textbf{Low-level projection:} \quad Z_L = [f^2_\omega \circ f^1_\theta(Y_1) \dots f^2_\omega \circ f^1_\theta(Y_T)]$, \\
$\textbf{High-level projection:} \quad Z_H = \sum_{i=1}^T f^{3}_{\omega'_i} \circ f^{1}_{\theta}(Y_i)$. 

The hierarchical autocorrelation matrices become:
\begin{gather}    
    R_L = \mathbb{E}[Z^L (Z^{L})^\top], \quad R_H = \mathbb{E}[Z^H (Z^{H})^\top], \\ 
    P_{LH} = \mathbb{E}[Z^L (Z^{H})^\top], \quad R_{LH} = \begin{bmatrix}
R_L & P_{LH} \\
P_{LH}^\top & R_H
\end{bmatrix}.
\end{gather}

The training objective minimises:
\begin{equation}
\mathcal{L}_{HFMCA} = \log \det R_{LH} - \log \det R_L - \log \det R_H.
\end{equation}




Our augmentation strategy employs graph-specific transformations including random walk sampling, node dropping, feature masking, and edge removal. After pretraining, projection heads are discarded and only the backbone encoder $f^1_\theta$ is retained for downstream classification tasks.

\subsection{Graph Transformer Architecture}

We employ a Graph Transformer backbone based on the GPS architecture \citep{graphgps}, which combines local neighbourhood information via message-passing with global attention mechanisms. The architecture incorporates Random Walk Positional Encodings \cite{rwe} to preserve structural relationships between brain regions. After processing through multiple transformer layers, node embeddings are aggregated using global mean pooling to produce graph-level representations.


\begin{table}[t]
\centering
\begin{tabular}{lccc}
\toprule
Model & \makecell{REST\\(MDD)} & \makecell{REST\\(Sex)} & \makecell{ABIDE} \\
\midrule
\makecell{Majority\\class} & 51.6 $\pm$ 0.0 & 61.0 $\pm$ 0.0 & 53.6 $\pm$ 0.0 \\
\hdashline
\addlinespace
Baseline$_{F}$ & 55.5 $\pm$ 1.0 & 64.8 $\pm$ 1.1 & 53.5 $\pm$ 1.6 \\
VICReg$_{F}$ & 56.6 $\pm$ 1.0 & 65.4 $\pm$ 0.5 & 53.5 $\pm$ 1.3 \\
BT$_{F}$ & \textbf{57.3 $\pm$ 0.6} & 64.7 $\pm$ 0.7 & 53.5 $\pm$ 1.7 \\
SimCLR$_{F}$ & 55.9 $\pm$ 0.9 & \underline{65.9 $\pm$ 0.7} & \underline{54.0 $\pm$ 1.0} \\
\addlinespace
\hdashline
\addlinespace
HFMCA$_{F}$ & \textbf{57.4 $\pm$ 0.9} & \textbf{66.6 $\pm$ 1.3} & \textbf{54.7 $\pm$ 1.5} \\
\bottomrule
\end{tabular}
\caption{Classification accuracy (\%) with frozen encoders on datasets seen during pretraining.}
\label{tab:cls_freeze_seen}
\end{table}

\begin{table}[t]
\centering
\begin{tabular}{lccc}
\toprule
Model & \makecell{REST\\(MDD)} & \makecell{REST\\(Sex)} & \makecell{ABIDE} \\
\midrule
\makecell{Majority\\class} & 51.6 $\pm$ 0.0 & 61.0 $\pm$ 0.0 & 53.6 $\pm$ 0.0 \\
\hdashline
\addlinespace
\makecell{Baseline} & 58.8 $\pm$ 1.2 & \underline{68.3 $\pm$ 0.4} & 54.9 $\pm$ 1.8 \\
\makecell{VICReg} & 58.7 $\pm$ 1.0 & 65.5 $\pm$ 1.4 & 52.9 $\pm$ 1.7 \\
\makecell{BT} & \textbf{59.9 $\pm$ 1.4} & 65.2 $\pm$ 1.5 & 53.5 $\pm$ 1.0 \\
\makecell{SimCLR} & 59.3 $\pm$ 1.2 & \underline{68.4 $\pm$ 0.9} & \textbf{56.3 $\pm$ 1.3} \\
\addlinespace
\hdashline
\addlinespace
\makecell{HFMCA} & \textbf{59.8 $\pm$ 1.2} & \textbf{70.3 $\pm$ 1.0} & \textbf{56.1 $\pm$ 1.6} \\
\bottomrule
\end{tabular}
\caption{Classification accuracy (\%) with unfrozen encoders on previously seen datasets.}
\label{tab:cls_seen}
\end{table}

\section{Experiments}
\subsection{Datasets and Tasks}

We evaluate our approach on five neuroimaging datasets representing diverse classification challenges:\\
\textbf{REST} \citep{restmdd}: Major Depressive Disorder and sex classification (1642 subjects; 51.6\% Healthy/MDD ratio; 61.0\% Male/Female ratio)  \\
\textbf{ABIDE} \citep{abide}: Autism Spectrum Disorder classification (866 subjects; 53.6\% Healthy/ASD ratio) \\
\textbf{BSNIP} \citep{bsnip}: Schizophrenia (SZ) and Bipolar Disorder with Psychosis (BP) (1464 subjects; 43.7\%/34.2\%/22.1\% Healthy/SZ/BP ratio) \\
\textbf{AOMIC} \citep{aomic}: Sex classification (881 subjects; 51.9\% Male/
Female ratio) \\
\textbf{HCP} \citep{hcp}: Sex classification (443 subjects; 55.5\% Male/
Female ratio)
Models are pretrained on REST and ABIDE datasets (2005 subjects total), then evaluated on all datasets including previously unseen ones (BSNIP, AOMIC, HCP).

\subsection{Training Protocol and Evaluation}

The GPS encoder is equipped with two projection heads required for HFMCA training. We train for 200 epochs using Adam optimiser with learning rate $10^{-3}$, weight decay $10^{-6}$, and batch size 256. The model is pretrained on a combined REST and ABIDE dataset, comprising approximately 2500 samples. After pretraining, projection heads are discarded and only the backbone encoder is retained for downstream evaluation.

We compare HFMCA against established self-supervised baselines: SimCLR \citep{simclr}, Barlow Twins \citep{barlowtwins}, VICReg \citep{vicreg}, and a randomly initialised Baseline. Following standard practice \citep{misra,vicreg}, we evaluate both frozen and unfrozen scenarios using nested 5-fold cross-validation with 10 independent runs.

\subsection{Quality of Learned Representations}


To assess the quality of the learned representations, we attach a trainable linear classifier on top of the frozen graph encoder embeddings and evaluate performance on multiple classification tasks (Table \ref{tab:cls_freeze_seen}). We also unfreeze the encoder and fine-tune it on the same downstream tasks (Table \ref{tab:cls_seen}). HFMCA achieves competitive results across all benchmarks, demonstrating more consistent performance across runs.

\begin{table}
\centering
\begin{tabular}{lccc}
\toprule
Model & \makecell{BSNIP} & \makecell{AOMIC\\(Sex)} & \makecell{HCP\\(Sex)} \\
\midrule
\makecell{Majority\\class} & 43.7 $\pm$ 0.0 & 51.9 $\pm$ 0.0 & 55.5 $\pm$ 0.0 \\
\hdashline
\addlinespace
Baseline$_{F}$ & 47.7 $\pm$ 0.9 & 56.9 $\pm$ 1.3 & 63.0 $\pm$ 2.4 \\
VICReg$_{F}$ & 46.4 $\pm$ 0.9 & \underline{60.0 $\pm$ 1.5} & \underline{65.7 $\pm$ 1.5} \\
BT$_{F}$ & 46.2 $\pm$ 1.2 & \textbf{60.9 $\pm$ 0.8} & \underline{66.0 $\pm$ 1.5} \\
SimCLR$_{F}$ & \underline{47.9 $\pm$ 0.5} & 59.2 $\pm$ 1.5 & \textbf{66.6 $\pm$ 1.5} \\
\addlinespace
\hdashline
\addlinespace
HFMCA$_{F}$ & \textbf{48.7 $\pm$ 0.8} & 59.4 $\pm$ 1.1 & \underline{66.0 $\pm$ 1.9} \\
\bottomrule
\end{tabular}
\caption{Classification accuracy (\%) with frozen encoders on previously unseen datasets. }
\label{tab:cls_freeze_zeroshot}
\end{table}

\begin{table}
\centering
\begin{tabular}{lccc}
\toprule
Model & \makecell{BSNIP} & \makecell{AOMIC\\(Sex)} & \makecell{HCP\\(Sex)} \\
\midrule
\makecell{Majority\\class} & 43.7 $\pm$ 0.0 & 51.9 $\pm$ 0.0 & 55.5 $\pm$ 0.0 \\
\hdashline
\addlinespace
\makecell{Baseline} & 47.8 $\pm$ 0.8 & 62.5 $\pm$ 1.8 & \textbf{71.1 $\pm$ 2.2} \\
\makecell{VICReg} & 47.1 $\pm$ 1.0 & 60.5 $\pm$ 1.7 & 66.0 $\pm$ 3.2 \\
\makecell{BT} & 47.2 $\pm$ 1.2 & 60.3 $\pm$ 1.4 & 66.1 $\pm$ 1.1 \\
\makecell{SimCLR} & \textbf{48.7 $\pm$ 0.6} & \underline{63.7 $\pm$ 1.8} & 68.2 $\pm$ 3.1 \\
\addlinespace
\hdashline
\addlinespace
\makecell{HFMCA} & \underline{48.4 $\pm$ 0.8} & \textbf{64.6 $\pm$ 1.4} & \underline{70.2 $\pm$ 0.6} \\
\bottomrule
\end{tabular}
\caption{Zero-shot classification accuracy (\%) with unfrozen encoders.}
\label{tab:cls_zeroshot}
\end{table}

\subsection{Transfer Learning Evaluation}

To evaluate the transferability of the HFMCA encoder, we fine-tune both frozen and unfrozen variants with attached linear heads on tasks outside the pretraining datasets. The results (Tables \ref{tab:cls_freeze_zeroshot} and \ref{tab:cls_zeroshot}) show that HFMCA consistently outperforms random initialization (\textit{Baseline}) and remains competitive with other methods. Notably, it achieves more stable performance on average, exhibiting lower variance across experimental runs.

\subsection{Scaling Laws}

\begin{figure}[t]
\begin{minipage}{1.0\linewidth}
  \centering
  \centerline{\includegraphics[width=8.5cm]{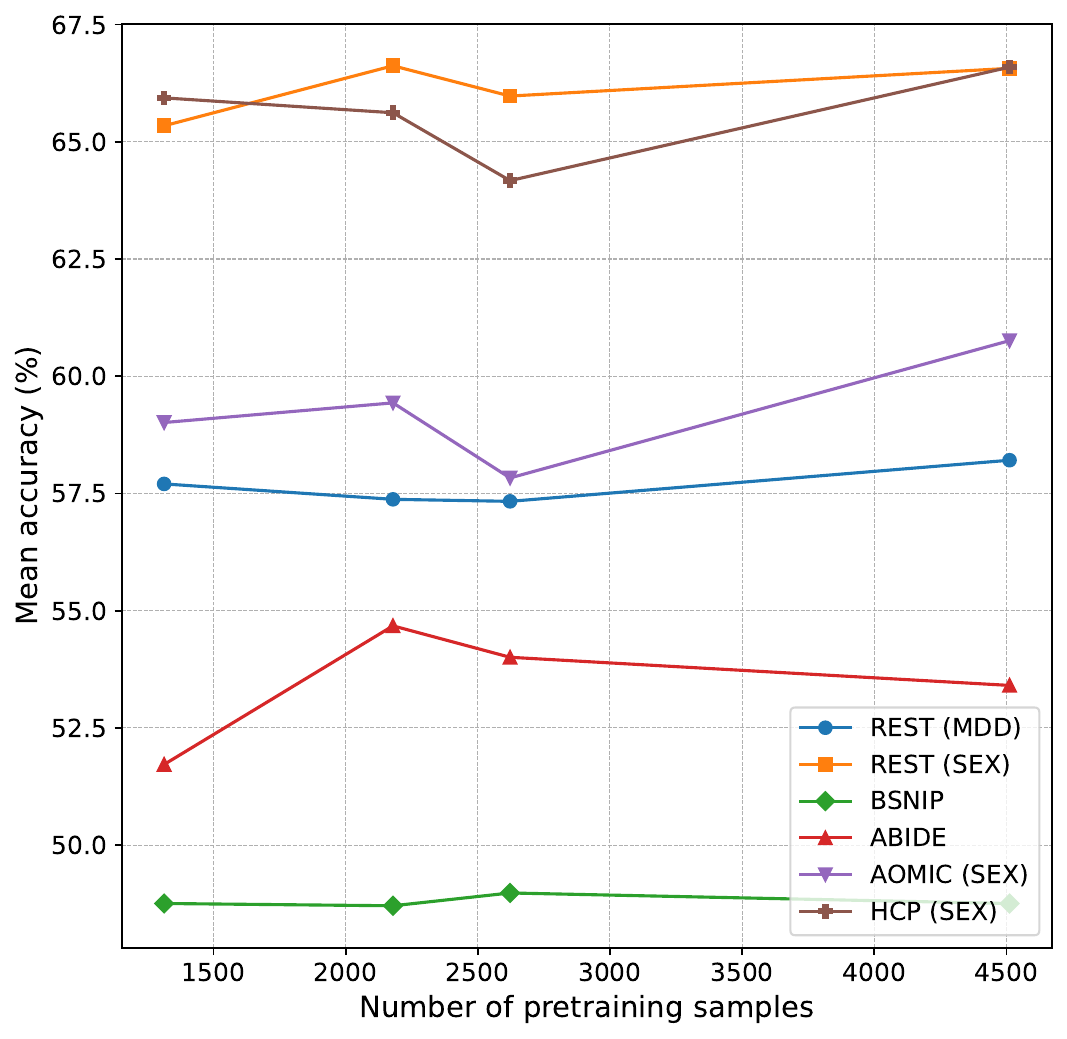}}
\end{minipage}
  \caption{Encoders were trained with HFMCA using varying amounts of pretraining data and fine-tuned with linear heads. No clear linear relationship was observed between the amount of pretraining data and downstream performance, which aligns with findings from recent studies.}
  \label{fig:scaling}
\end{figure}

To investigate whether pretraining performance scales with dataset size, we pretrained HFMCA encoders on progressively larger datasets:
\textbf{REST} (1313 subjects), \textbf{REST + ABIDE} (2179 subjects), \textbf{REST + ABIDE + HCP} (2622 subjects), and \textbf{REST + ABIDE + HCP + BSNIP} (4512 subjects). 
Each encoder was evaluated using frozen linear heads across all downstream tasks following the protocol from Section 4.3. Figure \ref{fig:scaling} shows that contrary to established scaling laws \citep{kaplan2020scaling,bansal2022data}, performance does not improve monotonically with increased data volume. 
Performance peaks with REST + ABIDE but declines when adding HCP and BSNIP datasets. 
This aligns with recent findings that naive scaling can produce negative transfer effects in graph foundation models \citep{scalefail1,scalefail2}.

\section{Conclusion}


We successfully extended HFMCA to graph-structured fMRI data, providing a theoretically principled approach to self-supervised representation learning. Our method achieves competitive performance across five neuroimaging datasets. The demonstrated transfer learning capabilities and stable training make HFMCA particularly suitable for neuroimaging applications. Future work should explore larger-scale datasets and investigate HFMCA as a component of foundational models for brain imaging. Even though the initial scaling law analysis suggests greater complexity compared to text and vision domains, the framework and transferability indicate important contributions toward generalizable computational models of brain function.

\renewcommand{\bibfont}{\small}
\bibliographystyle{IEEEbib}
\bibliography{main}

\begin{thebibliography}{10}

\bibitem{autism1}
C.~S. Monk, S.~J. Peltier, J.~L. Wiggins, S.-J. Weng, M.~Carrasco, S.~Risi, and C.~Lord,
\newblock ``Abnormalities of intrinsic functional connectivity in autism spectrum disorders,,''
\newblock {\em NeuroImage}, vol. 47, no. 2, pp. 764--772, 2009.

\bibitem{mdd1}
B.~de~Kwaasteniet, E.~Ruhe, M.~Caan, M.~Rive, S.~Olabarriaga, M.~Groefsema, L.~Heesink, G.~van Wingen, and D.~Denys,
\newblock ``Relation between structural and functional connectivity in major depressive disorder,''
\newblock {\em Biological Psychiatry}, vol. 74, no. 1, pp. 40--47, 2013.

\bibitem{schizophrenia1}
M.~Liang, Y.~Zhou, T.~Jiang, Z.~Liu, L.~Tian, H.~Liu, and Y.~Hao,
\newblock ``Widespread functional disconnectivity in schizophrenia with resting-state functional magnetic resonance imaging,''
\newblock {\em Neuroreport}, vol. 17, no. 2, pp. 209--213, Feb. 2006.

\bibitem{fmrissl1}
C.~Shi, Y.~Wang, Y.~Wu, S.~Chen, R.~Hu, M.~Zhang, B.~Qiu, and X.~Wang,
\newblock ``Self-supervised pretraining improves the performance of classification of task functional magnetic resonance imaging,''
\newblock {\em Frontiers in Neuroscience}, vol. 17, pp. 1199312, June 2023.

\bibitem{fmrissl2}
X.~Wang, Y.~Chu, Q.~Wang, L.~Cao, L.~Qiao, L.~Zhang, and M.~Liu,
\newblock ``Unsupervised contrastive graph learning for resting-state functional mri analysis and brain disorder detection,''
\newblock {\em Human Brain Mapping}, vol. 44, no. 17, pp. 5672--5692, Dec 2023,
\newblock Epub 2023 Sep 5.

\bibitem{fmrissl3}
J.~O. Caro et~al.,
\newblock ``Brain{LM}: A foundation model for brain activity recordings,''
\newblock in {\em ICLR}, 2024.

\bibitem{ss_transformers}
I.~Malkiel, G.~Rosenman, L.~Wolf, and T.~Hendler,
\newblock ``Self-supervised transformers for fmri representation,''
\newblock in {\em Proceedings of the 5th International Conference on Medical Imaging with Deep Learning}, 2022, pp. 895--913.

\bibitem{hfmca}
Bo~Hu, Yuheng Bu, and Jos{\'e}~C Pr{\'\i}ncipe,
\newblock ``Learning orthonormal features in self-supervised learning using functional maximal correlation,''
\newblock in {\em 2024 IEEE International Conference on Image Processing (ICIP)}, 2024, pp. 472--478.

\bibitem{simclr}
Ting Chen, Simon Kornblith, Mohammad Norouzi, and Geoffrey Hinton,
\newblock ``A simple framework for contrastive learning of visual representations,''
\newblock in {\em ICML}, 2020, pp. 1597--1607.

\bibitem{moco}
Kaiming He, Haoqi Fan, Yuxin Wu, Saining Xie, and Ross Girshick,
\newblock ``Momentum contrast for unsupervised visual representation learning,''
\newblock in {\em CVPR}, 2020, pp. 9729--9738.

\bibitem{barlowtwins}
Jure Zbontar, Li~Jing, Ishan Misra, Yann LeCun, and St{\'e}phane Deny,
\newblock ``Barlow twins: Self-supervised learning via redundancy reduction,''
\newblock in {\em ICML}, 2021, pp. 12310--12320.

\bibitem{vicreg}
Adrien Bardes, Jean Ponce, and Yann Lecun,
\newblock ``Vicreg: Variance-invariance-covariance regularization for self-supervised learning,''
\newblock in {\em ICLR}, 2022.

\bibitem{hfmca2}
S.~Ma, B.~Hu, T.~Jia, A.~K. Clarke, B.~Zicher, A.~H. Caillet, D.~Farina, and J.~C. Principe,
\newblock ``Learning cortico-muscular dependence through orthonormal decomposition of density ratios,''
\newblock in {\em The Thirty-eighth Annual Conference on Neural Information Processing Systems}, 2024.

\bibitem{aal116}
N.~Tzourio-Mazoyer et~al.,
\newblock ``Automated anatomical labeling of activations in spm using a macroscopic anatomical parcellation of the {MNI} {MRI} single-subject brain,''
\newblock {\em NeuroImage}, vol. 15, no. 1, pp. 273--289, 2002.

\bibitem{graphgps}
Ladislav Ramp{\'a}{\v{s}}ek, Michael Galkin, Vijay~Prakash Dwivedi, Anh~Tuan Luu, Guy Wolf, and Dominique Beaini,
\newblock ``Recipe for a general, powerful, scalable graph transformer,''
\newblock in {\em NeurIPS}, 2022, vol.~35, pp. 14501--14515.

\bibitem{rwe}
Vijay~Prakash Dwivedi, Anh~Tuan Luu, Thomas Laurent, Yoshua Bengio, and Xavier Bresson,
\newblock ``Graph neural networks with learnable structural and positional representations,''
\newblock in {\em International Conference on Learning Representations}, 2022.

\bibitem{restmdd}
X.~Chen et~al.,
\newblock ``The direct consortium and the rest-meta-mdd project: towards neuroimaging biomarkers of major depressive disorder,''
\newblock {\em Psychoradiology}, vol. 2, no. 1, pp. 32--42, 2022.

\bibitem{abide}
C.~Craddock, Y.~Benhajali, C.~Chu, F.~Chouinard, A.~Evans, A.~Jakab, B.~Khundrakpam, J.~D. Lewis, Q.~Li, M.~Milham, C.~Yan, and P.~Bellec,
\newblock ``The neuro bureau preprocessing initiative: open sharing of preprocessed neuroimaging data and derivatives,''
\newblock {\em Neuroinformatics}, 2013.

\bibitem{bsnip}
C.~A. Tamminga, G.~Pearlson, M.~Keshavan, J.~Sweeney, B.~Clementz, and G.~Thaker,
\newblock ``Bipolar and schizophrenia network for intermediate phenotypes: Outcomes across the psychosis continuum,''
\newblock {\em Schizophrenia Bulletin}, vol. 40, pp. S131--S137, 02 2014.

\bibitem{aomic}
L.~Snoek, M.~M. van~der Miesen, T.~Beemsterboer, A.~Van~Der Leij, A.~Eigenhuis, and H.~S. Scholte,
\newblock ``The amsterdam open mri collection, a set of multimodal mri datasets for individual difference analyses,''
\newblock {\em Scientific Data}, vol. 8, no. 1, pp. 1--23, 2021.

\bibitem{hcp}
D.~C.~Van Essen et~al.,
\newblock ``The human connectome project: A data acquisition perspective,''
\newblock {\em NeuroImage}, vol. 62, no. 4, pp. 2222--2231, 2012,
\newblock Connectivity.

\bibitem{misra}
Ishan Misra and Laurens van~der Maaten,
\newblock ``Self-supervised learning of pretext-invariant representations,''
\newblock in {\em CVPR}, 2020, pp. 6707--6717.

\bibitem{kaplan2020scaling}
J.~Kaplan, S.~McCandlish, T.~Henighan, T.~B. Brown, B.~Chess, R.~Child, S.~Gray, A.~Radford, J.~Wu, and D.~Amodei,
\newblock ``Scaling laws for neural language models,''
\newblock {\em arXiv preprint arXiv:2001.08361}, 2020.

\bibitem{bansal2022data}
Y.~Bansal et~al.,
\newblock ``Data scaling laws in nmt: The effect of noise and architecture,''
\newblock in {\em ICML}, 2022, pp. 1466--1482.

\bibitem{scalefail1}
Weihua Hu, Bowen Liu, Joseph Gomes, Marinka Zitnik, Percy Liang, Vijay Pande, and Jure Leskovec,
\newblock ``Strategies for pre-training graph neural networks,''
\newblock in {\em ICLR}, 2020.

\bibitem{scalefail2}
Yuxuan Cao et~al.,
\newblock ``When to pre-train graph neural networks? from data generation perspective!,''
\newblock in {\em Proceedings of the 29th ACM SIGKDD Conference on Knowledge Discovery and Data Mining}, 2023, pp. 142--153.

\end{thebibliography}

\end{document}